\documentclass[pra,showpacs,twocolumn]{revtex4}
\usepackage{amsmath,amsfonts,graphicx}
\usepackage[amssymb]{SIunits}
\usepackage{color}

\newcommand{\ket}[1]{|#1\rangle}
\newcommand{\bra}[1]{\langle#1|}
\newcommand{\bracket}[2]{\langle#1|#2\rangle}
\renewcommand{\vec}{\textbf}
\DeclareMathOperator{\id}{\openone}

\begin{document}

\bibliographystyle{apsrev}

\title{Helicity correlations of vector bosons}

\author{Pawe{\l}{} Caban}\email{P.Caban@merlin.phys.uni.lodz.pl}

\affiliation{Department of Theoretical Physics, University of Lodz\\
Pomorska 149/153, 90-236 {\L}{\'o}d{\'z}, Poland}

\date{\today}

\begin{abstract}
We calculate the helicity and polarization correlation functions in
the Einstein-Podolsky-Rosen-type experiments with relativistic vector
bosons. We show that the linear polarization correlation function in
the appriopriately chosen state in the massless limit is the same as
the correlation function in the scalar two-photon state. We show also
that the polarization correlation function violate the
Clauser-Horne-Shimony-Holt inequality and that the degree of this
violation can increase with the particle momentum.
\end{abstract}


\maketitle

\section{Introduction}


Starting from Czachor's paper \cite{Czachor1997_1},
Einstein-Podolsky-Rosen (EPR) correlations and other quantum
information primitives in the relativistic context have been widely
discussed
\cite{ALH2002, ALHK2003,ALMH2003, AM2002, BT2005, CR2003_Wigner,
CR2005, CR2006,CW2003, Czachor2005, GA2002, GBA2003, GKM2004,
Harshman2005, JSS2005, JSS2006, KM2003,KS2005, LD2003, LD2004,
LMS2005, LPT2003, LY2004, PST2002, PST2005, PT2003_2, PT2003_3,
PT2004_1, RS2002, SL2004, TU2003_1, YWYNMX2004, Caban2007_photons, CRW2008}.
For massive particles mostly the spin degrees of
freedom are considered. However, the definition of the spin operator
for the relativistic particle is not unique and different operators
have been used by different authors
\cite{Czachor1997_1,Terno2003,CW2003,LY2004,Czachor2005,CR2005}. 

On the other hand the 
helicity of the massive particle can be defined unambiguously and
recent papers \cite{HSZ2007_1,HSZ2007_2,SHZ2008} shows that helicity
entanglement of fermion pair 
behaves in a different way than spin entanglement.
Moreover, the carrier space of the irreducible, unitary representation
of the Poincar\'e group for photons is spanned by helicity
eingenstates. 
Furthermore, one can expect that the photon case can be received as a
massless limit of the spin 1 boson case. 
In the recent paper \cite{CRW2008} the
spin correlation function for the pair of vector bosons in the
covariant framework have been discussed and
the very surprising behavior of this function has been reported.  
In the center-of-mass frame for the definite 
configuration of the particles momenta and directions of the spin
projection measurements this correlation function still depends on
the value of the particle momentum and for some
configurations this dependence is not monotonic. In other words, for
fixed spin measurement directions and particle momenta directions,
the spin correlation function can have an extremum.

Therefore, it is very interesting to consider helicity correlation
function for the boson pair and the massless limit of this function. 
In this paper we calculate ordinary helicity correlation function of
the boson pair in certain scalar states. However, for photons usually
the polarization correlations are considered. Therefore, for massive
vector bosons we also define polarization states and calculate
polarization correlation function. In particular we identify the
two-particle scalar state which posseses a proper massless limit,
calculate polarization correlation function in this state, and show
that the massless limit of this function is the same as the
correlation function in the scalar two-photon state. We discuss also
the violation of the Clauser-Horne-Shimony-Holt (CHSH) inequality by
the polarization correlation function. We show that the degree of
violation of the CHSH inequality in the center of mass frame depends
on the particle momenta. What is very interesting, we find that the
degree of violation of the CHSH inequality can increase with the
particle momentum.

In Sec.~\ref{sec:representations} we recall basic facts concerning the
massive spin 1 representations of the Poincar\'e group in the helicity
basis. Section~\ref{sec:covariant_states} is devoted to the discussion
of states transforming covariantly with respect to the Poincar\'e
group action. In Sec.~\ref{sec:helicity_correlations} we calculate
explicitly helicity correlations of the boson pair in certain scalar
states. In Sec.~\ref{sec:polarization_states} we define states
corresponding to the longitudal and transversal polarization of vector
particle. In the next section we calculate polarization correlation
function for the boson pair; we discuss also properties of this
function and its massless limit. In this section we consider also 
the violation of Bell-type inequalities.
The last section contains concluding remarks. 

In the paper we use the natural units $\hbar=c=1$, the metric tensor
$\eta^{\mu\nu}=\textrm{diag}(1,-1,-1,-1)$, and the antisymmetric
tensor $\epsilon^{\alpha\beta\mu\nu}$ with $\epsilon^{0123}=1$.

\section{Representations of the Poincar\'e  group in the helicity basis}

\label{sec:representations}
In this section we recall the basic facts
concerning the spin-1 representation of the Poincar\'e group in the
helicity basis.
Let $\mathcal{H}$ be the carrier space of the
irreducible massive spin-1 representation of the Poincar\'e group. 
The spin basis of the space $\mathcal{H}$ we denote by 
$\{\ket{p,\sigma}\}_{\sigma=0,\pm1}$ while the helicity basis by
$\{\ket{p,\lambda}\}_{\lambda=0,\pm1}$. Vectors $\ket{p,\sigma}$
are eigenvectors of the four-momentum operators
 \begin{equation}
    \label{eq:fourmomentum_action}
    \hat{P}^{\mu}\ket{p,\sigma}=p^{\mu}\ket{p,\sigma},
 \end{equation}
where $p^2=m^2$ and $m$ is the mass of the particle.
We use the Lorentz-covariant normalization of the spin basis vectors
 \begin{equation}
    \label{eq:spin_basis_normalization}
    \bracket{p,\sigma}{p^\prime,\sigma^\prime} = 
    2 p^0 \delta^3(\vec{p}-\vec{p}^{\prime}) \delta_{\sigma\sigma^\prime}.
 \end{equation}
The vectors $\ket{p,\sigma}$ can be generated from standard vector
$\ket{\tilde{p},\sigma}$, where $\tilde{p}=m(1,0,0,0)$. We have
$\ket{p,\sigma}=U(L_p)\ket{\tilde{p},\sigma}$, where Lorentz boost
$L_p$ is defined by relations $p=L_p\tilde{p}$,
$L_{\tilde{p}}=\id$, and its explicit form is
 \begin{equation}
  \label{eq:explicit_Lp}
  L_p=\left(\begin{array}{c|c}
  \tfrac{p^0}{m} & \tfrac{\vec{p}^T}{m}\\
  \hline
  \tfrac{\vec{p}}{m} & \id+\tfrac{\vec{p}\otimes\vec{p}^T}{m(m+p^0)}
  \end{array}\right),
 \end{equation}
where $p^0=\sqrt{m^2+\vec{p}^2}$.

The standard Wigner induction procedure gives
 \begin{equation}
    \label{eq:representation_spin_basis}
    U(\Lambda)\ket{p,\sigma} = 
    \mathcal{D}_{\sigma^\prime\sigma}(R(\Lambda,p))
    \ket{\Lambda p,\sigma^\prime},
 \end{equation}
where the Wigner rotation $R(\Lambda,p)$ is defined as
$R(\Lambda,p)=L_{\Lambda p}^{-1}\Lambda L_p$ and $\mathcal{D}(R)$ is
the standard spin-1 representation of the rotation group. 
$3\times3$ matrix representations of the rotation group,
$\mathcal{D}(R)$ and $R$, are unitary equivalent, i.e.
for every rotation $R$
 \begin{equation}
   \label{eq:equivalent}
   \mathcal{D}(R)=VRV^{\dag},
 \end{equation}
where
 \begin{equation}
   \label{eq:matrix_V}
   V = \frac{1}{\sqrt{2}}\left(
      \begin{array}{ccc}
        -1 & i & 0 \\
        0 & 0 & \sqrt{2} \\
        1 & i & 0 \\
      \end{array}
    \right)
 \end{equation}
(see \cite{CRW2008}).

The helicity basis vectors $\ket{p,\lambda}$ are connected with the
spin basis vectors $\ket{p,\sigma}$ via the relation
\cite{cab_Weinberg1964}: 
 \begin{equation}
 \ket{p,\lambda} = \mathcal{D}_{\sigma\lambda}(R_{p})
 \ket{p,\sigma}, 
 \label{eq:helicity_with_spin}
 \end{equation}
where $R_{p}$ denotes the rotation which rotates the $z$-axis on
the direction of the vector $\vec{p}$, i.e.:
 \begin{equation}
 R_{p} \begin{pmatrix} 0\\0\\1 \end{pmatrix} = 
 \vec{n}_{\vec{p}} \equiv \frac{\vec{p}}{|\vec{p}|}.
 \label{eq:rotation_Rp}
 \end{equation}
The most general form of the rotation $R_{p}$ is
 \begin{equation}
 R_{p} = \left( 
 \vec{a}_{\vec{p}} \Big| \vec{n}_{\vec{p}} \times
 \vec{a}_{\vec{p}} \Big| \vec{n}_{\vec{p}} 
 \right),
 \label{cab:rotation_R_n_general}
 \end{equation}
where $|\vec{a}_{\vec{p}}|=1$, $\vec{a}_{\vec{p}}\perp
\vec{n}_{\vec{p}}$ and we treat vectors in
(\ref{cab:rotation_R_n_general}) as column matrices.
Vectors $\ket{p,\lambda}$ are eigenvectors of the four-momentum
operators 
 \begin{equation}
 \hat{P}^{\mu}\ket{p,\lambda}=p^{\mu}\ket{p,\lambda},
 \end{equation}
and Eqs.~(\ref{eq:spin_basis_normalization}), 
(\ref{eq:helicity_with_spin}) 
imply the Lorentz-covariant normalization of the helicity basis
 \begin{equation}
    \label{eq:helicity_basis_normalization}
    \bracket{p,\lambda}{p^\prime,\lambda^\prime} = 
    2 p^0 \delta^3(\vec{p}-\vec{p}^{\prime}) \delta_{\lambda\lambda^\prime}.
 \end{equation}

Vectors $\ket{p,\lambda}$ are eigenvectors of the helicity operator 
 \begin{equation}
 \frac{\vec{J}\cdot\vec{P}}{|\vec{P}|} \ket{p,\lambda} = \lambda
 \ket{p,\lambda},
 \label{eq:helity_operator_action}
 \end{equation}
where $\vec{J}=(J_1,J_2,J_3)$,
$J_i=\frac{1}{2}\varepsilon_{ijk}J_{ij}$ and 
$J_{\mu\nu}$ denote the generators of the Lorentz group such
that $U(\Lambda)=\exp(i\omega^{\mu\nu}J_{\mu\nu})$. Let us note
that $\vec{J}\cdot\vec{P} = W^0$, 
where the Pauli-Lubanski four-vector is defined as
 \begin{equation}
    \label{eq:def_Pauli_Lubanski}
    W^{\mu} = 
    \tfrac{1}{2} \epsilon^{\nu\alpha\beta\mu} P_{\nu}
    J_{\alpha\beta}. 
 \end{equation}
The relation (\ref{eq:helicity_with_spin}) can be inverted
 \begin{equation}
   \ket{p,\sigma} = \mathcal{D}_{\lambda\sigma}(R_{p}^{-1}) \ket{p,\lambda}.
   \label{eq:spin_with_helicity}
 \end{equation}
Now, from
Eqs.~(\ref{eq:representation_spin_basis}), 
(\ref{eq:helicity_with_spin}), and (\ref{eq:spin_with_helicity}) we
have 
 \begin{equation}
   U(\Lambda) \ket{p,\lambda} = \mathcal{D}_{\lambda^\prime\lambda}
   (R_{\Lambda p}^{-1} R(\Lambda,p) R_p) 
    \ket{\Lambda p,\lambda^\prime}. 
   \label{eq:representation_helicity_basis}
 \end{equation}

\section{Covariant states}

\label{sec:covariant_states}
Now, following \cite{CRW2008}, let us define states
 \begin{equation}
    \label{eq:covariant_state}
    \ket{(\mu,p)}=e^{\mu}_{\phantom{\mu}\sigma}(p)\ket{p,\sigma},
 \end{equation}
which transform covariantly
 \begin{equation}
    \label{eq:covariant_states_transformation}
    U(\Lambda)\ket{(\mu,p)} = (\Lambda^{-1})^{\mu}_{\phantom{\mu}\nu}
    \ket{(\nu,\Lambda p)}.
 \end{equation}
In Eq.~(\ref{eq:covariant_state}) $e_{\phantom{\mu}\sigma}^{\mu}(p)$
denotes the vector boson field amplitudes (see \cite{CRW2008} for the
details). Consistency of Eqs.~(\ref{eq:representation_spin_basis}),
(\ref{eq:covariant_state}), and
(\ref{eq:covariant_states_transformation}) is guaranteed by the 
Weinberg condition \cite{CRW2008} 
 \begin{equation}
   e^{\mu}_{\phantom{\mu}\sigma}(\Lambda p) =
    \Lambda^{\mu}_{\phantom{\mu}\nu} 
    e^{\nu}_{\phantom{\nu}\sigma^\prime}(p)
    \mathcal{D}(R(\Lambda,p))_{\sigma\sigma^\prime}.
   \label{eq:amplitudes_Weinberg}
 \end{equation}
The explicit form of the amplitudes $e_{\phantom{\mu}\sigma}^{\mu}(p)$
can be determined from Eq.~(\ref{eq:amplitudes_Weinberg}) (see \cite{CRW2008})
 \begin{equation}
    \label{eq:explicit_e(p)}
    e(p)=\left(\begin{array}{c}
   \tfrac{\vec{p}^T}{m}\\
   \hline
   \id+\tfrac{\vec{p}\otimes\vec{p}^T}{m(m+p^0)}
   \end{array}\right)V^{\mathrm{T}}.
 \end{equation}
Moreover, amplitudes (\ref{eq:explicit_e(p)}) fulfil the following conditions:
 \begin{subequations} 
 \label{seq:amplitudes}
   \begin{gather}
    p_\mu e^{\mu}_{\phantom{\mu}\sigma}(p) = 0,
   \label{eq:amplitudes_eq1} \\
    e^{*\mu}_{\phantom{*\mu}\sigma}(p) e_{\mu\sigma^\prime}(p) =
    - \delta_{\sigma\sigma^\prime},
   \label{eq:amplitudes_eq2} \\
    e^{\mu}_{\phantom{\mu}\sigma}(p) e_{\mu\sigma^\prime}(p) = 
    -(VV^{\mathrm{T}})_{\sigma\sigma^\prime},
   \label{eq:amplitudes_eq3} \\
    e^{*\mu}_{\phantom{*\mu}\sigma}(p)
    e^{\nu}_{\phantom{\nu}\sigma}(p) =
    - \eta^{\mu\nu}+\tfrac{p^{\mu}p^{\nu}}{m^2},
   \label{eq:amplitudes_eq4}
   \end{gather}
\end{subequations}
where $e(p)VV^{\mathrm{T}}=e^*(p)$, and $VV^{\mathrm{T}}=\left(
                                   \begin{array}{ccc}
                                     0 & 0 & -1 \\
                                     0 & 1 & 0 \\
                                     -1 & 0 & 0 \\
                                   \end{array}
                                 \right)$.

The covariant states (\ref{eq:covariant_state})
are normalized as follows [c.f.~Eq.~(\ref{eq:spin_basis_normalization})]
 \begin{equation}
    \label{eq:covariant_states_normaliztion}
    \bracket{(\mu, p^\prime)}{(\nu,p)} = 2p^0
    \delta^3(\vec{p}^\prime-\vec{p})
    e^{*\mu}_{\phantom{*\mu}\sigma}(p^\prime)
    e^{\nu}_{\phantom{\nu}\sigma}(p). 
 \end{equation}
Using helicity basis we have
 \begin{equation}
 \ket{(\mu,p)} = e^{\mu}_{\phantom{\mu}\sigma}(p) 
   {\mathcal D}_{\lambda\sigma}(R_{p}^{-1}) \ket{p,\lambda}
   \equiv E^{\mu}_{\phantom{\mu}\lambda}(p) \ket{p,\lambda}. 
 \label{eq:covariant_by_helicity} 
 \end{equation}
Equations (\ref{eq:equivalent}), (\ref{eq:explicit_e(p)}), and 
(\ref{seq:amplitudes}) imply 
 \begin{subequations} 
 \label{seq:amplitudes_E}
 \begin{gather}
   p_\mu E^{\mu}_{\phantom{\mu}\lambda}(p) = 0,
   \label{eq:orthogonal_E_p}\\
   E^{*\mu}_{\phantom{*\mu}\lambda}(p) E_{\mu\lambda^\prime}(p) =
   -\delta_{\lambda\lambda^\prime}, \\
   E^{\mu}_{\phantom{\mu}\lambda}(p) E_{\mu\lambda^\prime}(p) =
   -(VV^T)_{\lambda\lambda^\prime}, \\
   E^{*\mu}_{\phantom{*\mu}\lambda}(p) E^{\nu}_{\phantom{\nu}\lambda}(p) =
   -\eta^{\mu\nu} + \frac{p^\mu p^\nu}{m^2},
 \end{gather}
 \end{subequations}
and
 \begin{equation}
   E(p) = 
   \left(\begin{array}{c}
   \tfrac{\vec{p}^T}{m}\\
   \hline
   \id+\tfrac{\vec{p}\otimes\vec{p}^T}{m(m+p^0)}
   \end{array}\right) R_p V^{\mathrm{T}}.
   \label{eq:explicit_E(p)}
 \end{equation}
Thus, applying Eqs.~(\ref{eq:matrix_V}),
(\ref{cab:rotation_R_n_general}), and (\ref{eq:explicit_E(p)}) we have
explicitly 
 \begin{equation}
 E(p) = \left( \begin{array}{c|c|c}
 0 & \frac{\vec{p}^2}{m|\vec{p}|} & 0 \\ \hline
 \frac{-\vec{a}_{\vec{p}} + i
   (\vec{n}_{\vec{p}}\times\vec{a}_{\vec{p}})}{\sqrt{2}} &
 \frac{p^0\vec{p}}{m|\vec{p}|} &  
 \frac{\vec{a}_{\vec{p}} + i
   (\vec{n}_{\vec{p}}\times\vec{a}_{\vec{p}})}{\sqrt{2}}
 \end{array} \right),
 \label{eq:more_explicit_E(p)}
 \end{equation}
where the first column of the above matrix corresponds to $\lambda=1$,
the second one to 
$\lambda=0$, and the third one to $\lambda=-1$.
For futher convenience let us introduce the following notation
 \begin{equation}
 \ket{(k,\lambda);(p,\lambda^\prime)} \equiv \frac{1}{\sqrt{2}} 
 \big[ \ket{k,\lambda}\otimes\ket{p,\lambda^\prime} +
 \ket{p,\lambda^\prime}\otimes\ket{k,\lambda}  \big].
 \end{equation}
For $k\not=p$ Eq.~(\ref{eq:helicity_basis_normalization}) imply
 \begin{equation}
 \bracket{(k,\tau);(p,\tau^\prime)}{(k,\lambda);(p,\lambda^\prime)} =
 4 k^0 p^0 \big[ \delta^3(\vec{0}) \big]^2 
 \delta_{\tau\lambda} \delta_{\tau^\prime\lambda^\prime}.
 \label{eq:helicity_two_particle_normalization}
 \end{equation}

Finally, with help of one-particle covariant states
(\ref{eq:covariant_state}), we can define two-particle states which
posses defined transformation properties under Lorentz group
action. In particular, the most general scalar state with sharp
momenta has the following form: 
 \begin{equation}
 \ket{\chi(k,p)} = \alpha(k,p) (kp) \ket{\psi(k,p)} + \beta(k,p)
 \ket{\phi(k,p)}, 
 \label{eq:scalar_state_general}
 \end{equation}
where two independent scalar states, $\ket{\psi(k,p)}$ and
$\ket{\phi(k,p)}$, are given by
 \begin{align}
 \ket{\psi(k,p)} & = \eta_{\mu\nu} E^{\mu}_{\phantom{\mu}\lambda}(k)
 E^{\nu}_{\phantom{\nu}\lambda^\prime}(p)
 \ket{(k,\lambda);(p,\lambda^\prime)},  
 \label{eq:scalar_state_psi}\\
 \ket{\phi(k,p)} & = 
 \big[ p_\mu E^{\mu}_{\phantom{\mu}\lambda}(k) \big]
 \big[ k_\nu E^{\nu}_{\phantom{\nu}\lambda^\prime}(p) \big] 
 \ket{(k,\lambda);(p,\lambda^\prime)},
 \label{eq:scalar_state_phi}
 \end{align}
and $\alpha(k,p)$, $\beta(k,p)$ denote scalar functions
of four-momenta $k$ and $p$.
In the EPR-type experiments Alice and Bob measure correlation function
in the two-particle state with different momenta. Therefore, in the
rest of the paper we will assume that in the states
(\ref{eq:scalar_state_psi}), (\ref{eq:scalar_state_phi}) 
$k\not=p$. In this case it holds
 \begin{align}
 \bracket{\psi(k,p)}{\psi(k,p)} & = 4 k^0 p^0
  \big[\delta^3(\vec{0})\big]^2 
  \Big[ \frac{(kp)^2}{m^4} + 2 \Big], 
 \label{eq:normalization_scalar_psi} \\
 \bracket{\phi(k,p)}{\phi(k,p)} & = 4 k^0 p^0
  \big[\delta^3(\vec{0})\big]^2 m^4 
  \Big[\frac{(kp)^2}{m^4} -1 \Big]^2, 
 \label{eq:normalization_scalar_phi} \\
 \bracket{\phi(k,p)}{\psi(k,p)} & = 4 k^0 p^0
  \big[\delta^3(\vec{0})\big]^2 (kp)
  \Big[\frac{(kp)^2}{m^4} - 1 \Big]. 
 \label{eq:normalization_scalar_phi_psi} 
 \end{align}
With help of the above equations we can easily find the square of the
norm of the state $\ket{\chi(k,p)}$
 \begin{multline}
 \bracket{\chi(k,p)}{\chi(k,p)} = 4 k^0 p^0
 \big[\delta^3(\vec{0})\big]^2 \\ \times
 \Big\{ |\alpha|^2 (kp)^2 \Big[ \frac{(kp)^2}{m^4} + 2 \Big] 
 + |\beta|^2 m^4 \Big[\frac{(kp)^2}{m^4} -1 \Big]^2 +\\
 + (\alpha^* \beta + \alpha \beta^*) (kp)^2
   \Big[\frac{(kp)^2}{m^4} - 1 \Big] \Big\}.
 \label{eq:normalization_scalar_general}
 \end{multline}
In the discussion of the massless limit of the correlation function,
the following state will be of special interest
 \begin{equation}
 \ket{\xi(k,p)} = -(kp) \ket{\psi(k,p)} + \ket{\phi(k,p)}.
 \label{eq:scalar_state_ksi}
 \end{equation}
For the state (\ref{eq:scalar_state_ksi}) it holds 
 \begin{equation}
 \bracket{\xi(k,p)}{\xi(k,p)} = 4 k^0 p^0
 \big[\delta^3(\vec{0})\big]^2 \Big[ 2 (kp)^2 + m^4 \Big].
 \label{eq:normalization_scalar_ksi}
 \end{equation}

\section{Helicity correlations}

\label{sec:helicity_correlations}
Now, we will calculate helicity correlations in the EPR type
experiments. We assume that Alice and Bob share two-particle scalar
state defined in Eq.~(\ref{eq:scalar_state_general})
(with $k\not=p$) and Alice measures the helicity of the
particle with the four-momentum $k$ and Bob the helicity of the
particle with the four-momentum $p$. Therefore, Alice uses the
observable 
 \begin{equation}
 \Lambda_k = \hat{\lambda}_k\otimes\openone +
 \openone\otimes\hat{\lambda}_k,
 \end{equation}
where
 \begin{equation}
 \hat{\lambda}_k \ket{k,\lambda} = \lambda \ket{k,\lambda},
 \label{eq:lambda_k_1}
 \end{equation}
and
 \begin{equation}
 \hat{\lambda}_k \ket{p,\lambda} = 0,\quad k\not=p;
 \label{eq:lambda_k_2}
 \end{equation}
while Bob uses the observable
 \begin{equation}
 \Lambda_p = \hat{\lambda}_p\otimes\openone +
 \openone\otimes\hat{\lambda}_p.
 \end{equation}
$\hat{\lambda}_p$ fulfills relations analogous to
Eqs.~(\ref{eq:lambda_k_1}) and (\ref{eq:lambda_k_2}).
Helicity correlation function has the form
 \begin{equation}
 \mathcal{C}_{\textsf{helicity}}^{\chi(k,p)}(k,p) = 
 \frac{\bra{\chi(k,p)}\Lambda_p \Lambda_k
 \ket{\chi(k,p)}}{\bracket{\chi(k,p)}{\chi(k,p)}}.
 \label{eq:helicity_correlations_def}
 \end{equation}
Taking into account Eqs.~(\ref{eq:explicit_E(p)}), 
(\ref{eq:scalar_state_psi}), (\ref{eq:scalar_state_phi}), 
and (\ref{eq:helicity_two_particle_normalization}) after
strightforward calculation, we receive
\begin{subequations}
 \begin{align}
 \bra{\psi(k,p)}\Lambda_p \Lambda_k \ket{\psi(k,p)} & = 
 -8 k^0 p^0 \big[ \delta^3(\vec{0}) \big]^2 
 \frac{\vec{k}\cdot\vec{p}}{|\vec{k}||\vec{p}|},\\
 \bra{\phi(k,p)}\Lambda_p \Lambda_k \ket{\phi(k,p)} & = 0, 
 \label{eq:helicity_matrix_elements_phi_phi}\\
 \bra{\phi(k,p)}\Lambda_p \Lambda_k \ket{\psi(k,p)} & = 
 - k^0 p^0 \big[ \delta^3(\vec{0}) \big]^2
 \frac{(\vec{k}\times\vec{p})^2}{|\vec{k}||\vec{p}|}.
 \end{align}
\label{seq:helicity_matrix_elements}
\end{subequations}
Now, using Eqs.~(\ref{eq:scalar_state_general}),
(\ref{eq:normalization_scalar_general}), and
(\ref{seq:helicity_matrix_elements}) one can easily calculate
correlation function (\ref{eq:helicity_correlations_def}) for
arbitrary state $\ket{\chi(k,p)}$. We give here explicit formulas for
some special cases. First of all,
Eq.~(\ref{eq:helicity_matrix_elements_phi_phi}) imply that for the
state (\ref{eq:scalar_state_phi}) we have
  \begin{equation}
 \mathcal{C}_{\textsf{helicity}}^{\phi(k,p)}(k,p) = 0.
 \label{eq:helicity_correlations_phi}
 \end{equation}
For the state (\ref{eq:scalar_state_psi}) we have
 \begin{equation}
 \mathcal{C}_{\textsf{helicity}}^{\psi(k,p)}(k,p) = 
 - \frac{2}{\big[ 2 + \frac{(kp)^2}{m^4} \big]} 
 \frac{\vec{k}\vec{p}}{|\vec{k}| |\vec{p}|}.
 \label{eq:helicity_correlations}
 \end{equation}

It should be noted that the standard spin
correlation function in the EPR-type expiriment 
in which Alice and Bob share the state
(\ref{eq:scalar_state_psi}) and measure the spin components in 
directions $\vec{a}$ and $\vec{b}$, respectively, was calculated
in Ref.~\cite{CRW2008}.
This function has the following form \cite{CRW2008}: 
 \begin{multline}
 \mathcal{C}_{\textsf{spin}}^{\psi(k,p)}(k,\vec{a};p,\vec{b}) = 
 \frac{2}{m^2 \big[ 2 + \frac{(kp)^2}{m^4} \big]} 
 \Big\{ -(\vec{a}\cdot\vec{b})(kp) + \\
 - (\vec{a}\cdot\vec{p})(\vec{b}\cdot\vec{k}) 
 - \frac{(\vec{a}\cdot\vec{k})(\vec{b}\cdot\vec{p})
     (\vec{k}\cdot\vec{p})}{(m+k^0)(m+p^0)} + \\
 + \frac{k^0(\vec{a}\cdot\vec{p})(\vec{b}\cdot\vec{p})}{m+p^0}
 + \frac{p^0(\vec{a}\cdot\vec{k})(\vec{b}\cdot\vec{k})}{m+k^0}
 \Big\}.
 \label{eq:spin_correlations}
 \end{multline}
In \cite{CRW2008} the following relativistic spin operator 
 \begin{equation} 
   \label{eq:def_spin_operator}
   \vec{S} = \frac{1}{m} \left( \vec{W} - W^0
     \frac{\vec{P}}{P^0+m} \right),
 \end{equation}
has been used. For this operator
it holds $\vec{S}\cdot\vec{P} = W^0$. Therefore, we can write
 \begin{equation}
   \frac{\vec{S}\cdot\vec{P}}{|\vec{P}|} \ket{p,\lambda} = \lambda
   \ket{p,\lambda}.
   \label{eq:helity_operator_action_2}
 \end{equation}
Thus, one can expect that choosing 
directions of Alice's and Bob's measurements parallel to $\vec{k}$ and
$\vec{p}$, respectively, i.e.\ putting in Eq.~(\ref{eq:spin_correlations})
 \begin{equation}
 \vec{a} = \frac{\vec{k}}{|\vec{k}|}, \quad
 \vec{b} = \frac{\vec{p}}{|\vec{p}|},
 \end{equation}
one should receive Eq.~(\ref{eq:helicity_correlations}). Simple
calculation shows that this is really the case.  

Finally, helicity correlation function in the state
(\ref{eq:scalar_state_ksi}) has the form
 \begin{equation}
 \mathcal{C}_{\textsf{helicity}}^{\xi(k,p)}(k,p) = 
 \frac{-2(kp)}{2(kp)^2 + m^4}
 \frac{(kp) (\vec{k}\cdot\vec{p}) -
 (\vec{k}\times\vec{p})^2}{|\vec{k}| |\vec{p}|}.
 \label{eq:helicity_correlations_ksi}
 \end{equation}

\section{Polarization states}

\label{sec:polarization_states}
Now, we define vectors describing polarized states
 \begin{equation}
  \ket{\boldsymbol{\varepsilon}(p),p} = \varepsilon_\mu(p) \ket{(\mu,p)} = 
  \varepsilon_\mu(p) E^{\mu}_{\phantom{\mu}\lambda}(p) \ket{p,\lambda}.
  \label{eq:polarization_vectors_def}
 \end{equation}
It should be noted that without loss of generality we can assume
 \begin{equation}
 \varepsilon_\mu(p) p^\mu =0,
 \end{equation}
what is a consequence of Eq.~(\ref{eq:orthogonal_E_p}). Vectors
(\ref{eq:polarization_vectors_def}) are normalized as follows:
 \begin{equation}
   \bracket{\boldsymbol{\varepsilon}(p),p}{\boldsymbol{\varepsilon}^\prime(p^\prime),p^\prime}
   = -2 p^0 \delta^3(\vec{p}-\vec{p}^\prime)
   \varepsilon_{\phantom{*}\mu}^{*}(p) \varepsilon^{\prime\mu}(p).
   \label{eq:normalization_polarization_vectors}
 \end{equation}

\subsection{Longitudal polarization}

We say that vector (\ref{eq:polarization_vectors_def}) describes
longitudal polarization, when $\boldsymbol{\varepsilon}(p)||\vec{p}$. 
Therefore, using Eqs.~(\ref{eq:orthogonal_E_p}),
(\ref{eq:explicit_E(p)}), and (\ref{eq:matrix_V}), (\ref{eq:rotation_Rp})
one can easily show that 
 \begin{equation}
 \ket{\boldsymbol{\varepsilon}_{||}(p),p} = \ket{p,0},
 \label{eq:longitudal_vec}
 \end{equation}
and the corresponding polarization vector has the form
 \begin{equation}
  \boldsymbol{\varepsilon}_{||}(p)^\mu = 
  - \Big(\frac{|\vec{p}|}{m},\frac{p^0}{m|\vec{p}|}\vec{p}\Big).
 \end{equation}
In Eq.~(\ref{eq:longitudal_vec}) $\ket{p,0}$ denotes state with helicity 0.

\subsection{Transversal polarization}

Transversally polarized state is described by the vector
(\ref{eq:polarization_vectors_def}) with polarization vector
$\boldsymbol{\varepsilon}_\perp(p)$ orthogonal to the momentum
$\vec{p}$. This condition and Eqs.~(\ref{eq:orthogonal_E_p},
\ref{eq:explicit_E(p)}, \ref{eq:matrix_V}, \ref{eq:rotation_Rp}) imply
 \begin{equation}
   \ket{\boldsymbol{\varepsilon}_\perp(p),p} = \alpha \ket{p,1} + \beta
   \ket{p,-1}, 
   \label{eq:transversal_polarization_general_state}
 \end{equation}
where $|\alpha|^2+|\beta|^2=1$, $\ket{p,\pm1}$ are helicity states, and
 \begin{align}
    & \varepsilon_{\perp}(p)^0 = 0,\\
    & \boldsymbol{\varepsilon}_\perp(p) = 
   \frac{1}{\sqrt{2}} \Big[ (\alpha-\beta)\vec{a}_{\vec{p}} + 
   i (\alpha+\beta) (\vec{n}_{\vec{p}}\times \vec{a}_{\vec{p}}) \Big].
 \end{align}
States (\ref{eq:transversal_polarization_general_state}) are
normalized according to [compare
Eq.~(\ref{eq:normalization_polarization_vectors})] 
 \begin{equation}
 \bracket{\boldsymbol{\varepsilon}_\perp(p),p}%
{\boldsymbol{\varepsilon}_{\perp}^{\prime}(p^\prime),p^\prime} = 
 2 p^0 \delta^3(\vec{p}-\vec{p}^\prime)
 \boldsymbol{\varepsilon}_{\perp}^{*}(p)\cdot
 \boldsymbol{\varepsilon}_{\perp}^{\prime}(p).
 \end{equation}

\subsection{Circular and linear polarization}

Circularly polarized states we recive from
Eq.~(\ref{eq:transversal_polarization_general_state}) putting
$\alpha=1$, $\beta=0$ or $\alpha=0$, $\beta=1$
 \begin{align}
   & \ket{\boldsymbol{\varepsilon}_{\perp}^{+}(p),p} = \ket{p,1},\\
   & \ket{\boldsymbol{\varepsilon}_{\perp}^{-}(p),p} = \ket{p,-1}.
 \end{align}
Linearly polarized states we define in a standard way
 \begin{align}
   \ket{\boldsymbol{\varepsilon}_{\theta}(p),p}
   & = \frac{1}{\sqrt{2}} \big[ e^{i\theta}
   \ket{\boldsymbol{\varepsilon}_{\perp}^{+}(p),p}
   + e^{-i\theta} \ket{\boldsymbol{\varepsilon}_{\perp}^{-}(p),p}
   \big] \\
   & = \frac{1}{\sqrt{2}} \big[ e^{i\theta} \ket{p,1} + e^{-i\theta}
   \ket{p,-1} \big].
   \label{eq:linearly_polarized_def}
 \end{align}
Therefore the polarization vector for the linearly polarized state has
the form
 \begin{equation}
   \boldsymbol{\varepsilon}_{\theta}(p) = i \Big[ \sin{\theta}
   \vec{a}_{\vec{p}} + \cos{\theta} (\vec{n}_{\vec{p}} \times
   \vec{a}_{\vec{p}}) \Big].
 \label{eq:polarization_vector_explicit}
 \end{equation}
States (\ref{eq:linearly_polarized_def}) are normalized as follows
 \begin{equation}
   \bracket{\boldsymbol{\varepsilon}_{\theta}(p),
   p}{\boldsymbol{\varepsilon}_{\theta^\prime}(p^\prime),p^\prime} = 
   2 p^0 \delta^3(\vec{p}-\vec{p}^\prime) \cos{(\theta-\theta^\prime)}.
 \end{equation}

\section{Linear polarization correlations}

In this section we will consider EPR-type experiment in which Alice
and Bob measure linear polarization of vector particles. Precisely, 
we assume that Alice and Bob share two-particle scalar state,
particle with four-momentum $k$ flies to Alice, and particle with
four-momentum $p$ flies to Bob. Alice (Bob)
measures an observable which gives $+1$ when acting on boson with
four-momentum $k$ ($p$), polarized linearly under the angle $\theta$
($\tilde{\theta}$), and $-1$ for the similar boson polarized under the
angle $\theta_\perp = \theta + \pi/2$
($\tilde{\theta}_\perp$). Therefore, the observables used by Alice and
Bob have the following form:
 \begin{subequations}
 \begin{align}
 S^{\theta}_{k} & = \Pi^{\theta}_{k} - \Pi^{\theta_\perp}_{k},\\
 S^{\tilde{\theta}}_{p} & = \Pi^{\tilde{\theta}}_{p} -
 \Pi^{\tilde{\theta}_\perp}_{p}, 
 \end{align}
 \label{seq:observables}
 \end{subequations}
where
 \begin{multline}
 \Pi^{\theta}_{k} = \frac{1}{2k^0 \delta^3(\vec{0})} 
  \Big[
  \ket{\boldsymbol{\varepsilon}_{\theta}(k),k}
  \bra{\boldsymbol{\varepsilon}_{\theta}(k),k} \otimes \openone \\
  + \openone \otimes \ket{\boldsymbol{\varepsilon}_{\theta}(k),k}
  \bra{\boldsymbol{\varepsilon}_{\theta}(k),k}
  \Big],
 \end{multline}
and projectors $\Pi^{\theta_\perp}_{k}$, $\Pi^{\tilde{\theta}}_{p}$, 
$\Pi^{\tilde{\theta}_\perp}_{p}$ are
defined analogously to $\Pi^{\theta}_{k}$. 
It should be noted that observables (\ref{seq:observables}) commute
 \begin{equation}
 [S^{\theta}_{k},S^{\tilde{\theta}}_{p}] = 0,
 \label{eq:observables_commutator}
 \end{equation}
therefore the correlation function in the scalar state
(\ref{eq:scalar_state_general}) has the form 
 \begin{equation}
 \mathcal{C}^{\chi(k,p)}_{\textsf{polarization}}(k,\theta;p,\tilde{\theta}) = 
 \frac{\bra{\chi(k,p)} S^{\tilde{\theta}}_{p} S^{\theta}_{k}
 \ket{\chi(k,p)}}{\bracket{\chi(k,p)}{\chi(k,p)}}.
 \label{eq:correl_function_polarization_general}
 \end{equation}
The numerator of the right side of
Eq.~(\ref{eq:correl_function_polarization_general}) can be written as
[see Eqs.~(\ref{seq:observables})]
 \begin{align}
 \bra{\chi(k,p)} S^{\tilde{\theta}}_{p} S^{\theta}_{k}
 \ket{\chi(k,p)} & = \bra{\chi(k,p)}
 \Pi^{\tilde{\theta}}_{p} \Pi^{\theta}_{k} \ket{\chi(k,p)} \nonumber \\
  & + \bra{\chi(k,p)} \Pi^{\tilde{\theta}_\perp}_{p} 
   \Pi^{\theta_\perp}_{k} \ket{\chi(k,p)} \nonumber \\
  & - \bra{\chi(k,p)} \Pi^{\tilde{\theta}_\perp}_{p} 
   \Pi^{\theta}_{k} \ket{\chi(k,p)} \nonumber \\
  & - \bra{\chi(k,p)} \Pi^{\tilde{\theta}}_{p} 
   \Pi^{\theta_\perp}_{k} \ket{\chi(k,p)}.
 \label{eq:numerator_correl_func}
 \end{align}
It is enough to calculate explicitly only $\bra{\chi(k,p)}
\Pi^{\tilde{\theta}}_{p} \Pi^{\theta}_{k} \ket{\chi(k,p)}$, other terms on the right
side of Eq.~(\ref{eq:numerator_correl_func}) can be received by
appriopriate change of angles $\theta$ and/or $\tilde{\theta}$. 
Now, with help of Eqs.~(\ref{eq:covariant_by_helicity}),
(\ref{eq:explicit_E(p)}), (\ref{eq:scalar_state_psi}),
(\ref{eq:scalar_state_phi}), and
(\ref{eq:polarization_vector_explicit}) we find
\begin{subequations}
 \begin{align}
 \bra{\psi(k,p)} \Pi^{\tilde{\theta}}_{p} 
 \Pi^{\theta}_{k} \ket{\psi(k,p)} & = 4 k^0 p^0 
   \big[\delta^3(\vec{0}) \big]^2
   \big[ \boldsymbol{\varepsilon}_{\theta}(k) 
   \cdot\boldsymbol{\varepsilon}_{\tilde{\theta}}(p) \big]^2, 
 \label{eq:projectors_action_eq_1}\\
 \bra{\phi(k,p)} \Pi^{\tilde{\theta}}_{p} 
 \Pi^{\theta}_{k} \ket{\phi(k,p)} & = 4 k^0 p^0 
   \big[\delta^3(\vec{0}) \big]^2  
 \big[ \vec{p}\cdot\boldsymbol{\varepsilon}_{\theta}(k) \big]^2
 \nonumber \\
 & \times \big[ \vec{k}\cdot\boldsymbol{\varepsilon}_{\tilde{\theta}}(p)
 \big]^2, 
 \label{eq:projectors_action_eq_2}\\
 \bra{\phi(k,p)} \Pi^{\tilde{\theta}}_{p} 
 \Pi^{\theta}_{k} \ket{\psi(k,p)} & = - 4 k^0 p^0 
   \big[\delta^3(\vec{0}) \big]^2 
 \big[ \vec{p}\cdot\boldsymbol{\varepsilon}_{\theta}(k) \big]
 \nonumber \\
 & \times \big[ \vec{k}\cdot\boldsymbol{\varepsilon}_{\tilde{\theta}}(p) \big]
   \big[ \boldsymbol{\varepsilon}_{\theta}(k) 
   \cdot\boldsymbol{\varepsilon}_{\tilde{\theta}}(p) \big].
 \label{eq:projectors_action_eq_3}
 \end{align}
 \label{seq:projectors_action}
\end{subequations}
Using above formulas one can calculate explicitly
$\bra{\chi(k,p)} \Pi^{\tilde{\theta}}_{p} \Pi^{\theta}_{k}
\ket{\chi(k,p)}$ 
for arbitrary choice of functions $\alpha(k,p)$
and $\beta(k,p)$ in the state $\ket{\chi(k,p)}$ [see
Eq.~(\ref{eq:scalar_state_general})]. 
Thus, with help of Eqs.~(\ref{eq:numerator_correl_func}),
and (\ref{seq:projectors_action}) and the normalization
(\ref{eq:normalization_scalar_general}), the correlation function
(\ref{eq:correl_function_polarization_general}) can be easily
calculated for arbitrary scalar state 
$\ket{\chi(k,p)}$. However, the resulting general formula appears to
be rather long and we do not put it here. Instead of that we will
concentrate on some special cases. First of all, correlation functions
in the states $\ket{\psi(k,p)}$ [Eq.~(\ref{eq:scalar_state_psi})] and
$\ket{\phi(k,p)}$ [Eq.~(\ref{eq:scalar_state_phi})] have the following
form: 
 \begin{multline}
 \mathcal{C}^{\psi(k,p)}_{\textsf{polarization}}(k,\theta;p,\tilde{\theta}) = 
 \frac{1}{2+\frac{(kp)^2}{m^4}} \Big\{ 
 \Big[ 1 + \frac{(\vec{k}\cdot\vec{p})^2}{\vec{k}^2 \vec{p}^2} 
 \Big] \\
 \times
 \cos(2\theta) \cos(2\tilde{\theta}) + 
 2 \frac{\vec{k}\cdot\vec{p}}{|\vec{k}||\vec{p}|} \sin(2\theta)
 \sin(2\tilde{\theta}) 
 \Big\},
 \end{multline}
 \begin{equation}
 \mathcal{C}^{\phi(k,p)}_{\textsf{polarization}}(k,\theta;p,\tilde{\theta}) = 
 \frac{(\vec{k}\times\vec{p})^4 \cos(2\theta)
   \cos(2\tilde{\theta})}{m^4 \vec{k}^2 \vec{p}^2 \big[ 
   \frac{(kp)^2}{m^4} -1 \big]^2}.
 \end{equation}
In the center-of-mass frame these functions reduce to
 \begin{align}
 \mathcal{C}^{\psi(k,k^\pi)}_{\textsf{polarization}}
 (k,\theta;k^\pi,\tilde{\theta}) & =  
 \frac{2}{2+(2x+1)^2} \cos{2(\theta+\tilde{\theta})},
 \label{eq:correl_psi_helicity_cmf}\\
 \mathcal{C}^{\phi(k,k^\pi)}_{\textsf{polarization}}
 (k,\theta;k^\pi,\tilde{\theta}) & = 0,
 \end{align}
where $x=(\frac{\vec{k}}{m})^2$, $k^\pi=(k^0,-\vec{k})$. The function
(\ref{eq:correl_psi_helicity_cmf}) is monotonic and in the massless
limit ($x\to\infty$)
 \begin{equation}
 \lim\limits_{x\to\infty} \mathcal{C}^{\psi(k,k^\pi)}_{\textsf{polarization}}
 (k,\theta;k^\pi,\tilde{\theta}) = 0.
 \end{equation}

Now, let us consider the massless limit of the vector bosons
correlations. The general procedure of contracting massive
representation of Poincar\'e group to the massless one is very subtle
and we will not discuss it here. Let us only mention that one
of the main difficulties is connected to the fact that massive spin 1
particle posseses three helicity degrees of freedom while photons only
two. Thus, in the massless limit the transversal polarization and 
the longitudal polarization should transform separately. 

In the recent paper \cite{Caban2007_photons} the 
EPR-type experiments with photons in the Lorentz-covariant framework
have been discussed. In particular, the correlation function for the
pair of photons in the scalar state has been calculated. However, it
should be noted that there exists only one scalar state of two photons
with sharp momenta \cite{Caban2007_photons}, while there is a variety
of such states for two massive bosons [this variety corresponds to
different choices of $\alpha(k,p)$ and $\beta(k,p)$ in
Eq.~(\ref{eq:scalar_state_general})]. Thus, first of all we have to
identify the scalar two-boson state which in the massless limit goes
to the scalar two-photon state.
We claim that such a state has the form (\ref{eq:scalar_state_ksi}).
Indeed, in terms of helicity states we have
\begin{widetext}
 \begin{multline}
 \ket{\xi(k,p)} = \sum_{\lambda,\lambda^\prime=\pm1} 
 \Big\{ (kp) \big[ \vec{E}_{\lambda}(k) \cdot
 \vec{E}_{\lambda^\prime}(p) \big]
 + \big[ \vec{p}\cdot \vec{E}_{\lambda}(k) \big] 
 \big[ \vec{k}\cdot \vec{E}_{\lambda^\prime}(p) \big]\Big\}
 \ket{(k,\lambda);(p,\lambda^\prime)}
 + m^2 \frac{\vec{k}\cdot\vec{p}}{|\vec{k}||\vec{p}|}  \ket{(k,0);(p,0)}\\
 + m \sum_{\lambda=\pm1} \Big\{ \frac{k^0}{|\vec{p}|} 
 \big[ \vec{p} \cdot \vec{E}_{\lambda}(k) \big]
 \ket{(k,\lambda);(p,0)}
 + \frac{p^0}{|\vec{k}|} \big[ \vec{k} \cdot \vec{E}_{\lambda}(p) \big]
 \ket{(k,0);(p,\lambda)}
 \Big\},
 \label{eq:scalar_state_ksi_helicity}
 \end{multline}
where 
$\vec{E}_{\lambda}(k)=(E^{1}_{\lambda}(k),
E^{2}_{\lambda}(k),E^{3}_{\lambda}(k))$. 
Thus, taking into account the explicit form of $\vec{E}_\lambda(k)$
[Eq.~(\ref{eq:more_explicit_E(p)})], 
we see that the massless limit of
the state $\ket{\xi(k,p)}$ is well defined and that in this limit all
terms containing helicity 0 (longitudal polarization) vanish.
Using Eqs.~(\ref{seq:projectors_action}) we receive
 \begin{equation}
 \bra{\xi(k,p)} \Pi_{p}^{\tilde{\theta}} \Pi_{k}^{\theta}
 \ket{\xi(k,p)} = 
 4 k^0 p^0 \big[ \delta^3(\vec{0}) \big]^2 \Big\{
 (kp) \big[ \boldsymbol{\varepsilon}_{\theta}(k) \cdot
 \boldsymbol{\varepsilon}_{\tilde{\theta}}(p) \big] +
 \big[ \vec{k} \cdot \boldsymbol{\varepsilon}_{\tilde{\theta}}(p)
 \big] \big[
 \vec{p} \cdot \boldsymbol{\varepsilon}_{\theta}(k) \big]
 \Big\}^2,
 \end{equation}
and consequently correlation function in the state $\ket{\xi(k,p)}$
has the form 
 \begin{multline}
 \mathcal{C}^{\xi(k,p)}_{\textsf{polarization}}(k,\theta;p,\tilde{\theta}) =
 \frac{1}{2(kp)^2+m^4} \Big\{ 2(kp)^2 \Big[
 \cos(2\theta)\cos(2\tilde{\theta}) - \frac{k^0
   p^0}{|\vec{k}||\vec{p}|} \sin(2\theta)\sin(2\tilde{\theta}) \Big] \\
 + m^2(m^2+\vec{k}^2+\vec{p}^2) 
 \Big[ \frac{(kp)^2-2k^0p^0(kp)+
   m^2(m^2+\vec{k}^2+\vec{p}^2)}{\vec{k}^2\vec{p}^2}
 \cos(2\theta)\cos(2\tilde{\theta}) 
 +\frac{2(kp)}{|\vec{k}||\vec{p}|} \sin(2\theta)\sin(2\tilde{\theta})
 \Big] \Big\}.
 \label{eq:correl_ksi_polarization_general}
 \end{multline}
 \end{widetext}
It should be noted that in the massless limit the above correlation
function reduces to
 \begin{equation}
 \lim_{m\to0}
 \mathcal{C}^{\xi(k,p)}_{\textsf{polarization}}(k,\theta;p,\tilde{\theta}) = 
 \cos2(\theta+\tilde{\theta}),
 \end{equation}
which is just the correlation function for photons in the scalar state
obtained in \cite{Caban2007_photons}.

\begin{figure}
\includegraphics[width=1\columnwidth]{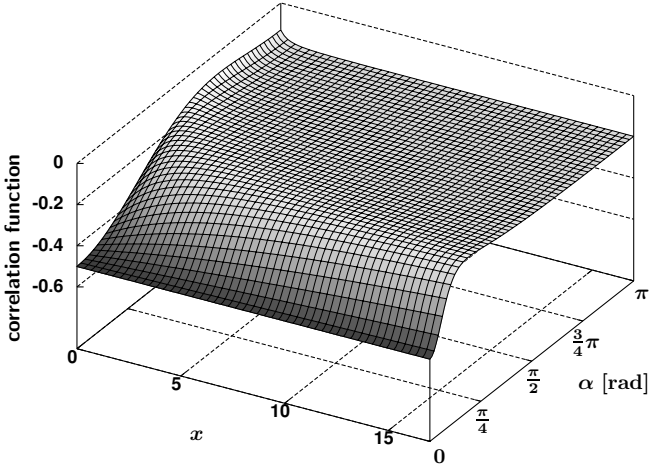}
\caption{The plot shows the dependence of the correlation function
  $\mathcal{C}^{\xi(k,p_k)}_{\textsf{polarization}} 
 (k,\theta;p_k,\tilde{\theta})$
 [Eq.~(\ref{eq:correlation_on_x_alpha})] on $x$ and $\alpha$ for
 $\theta = 5\pi/6$, $\tilde{\theta} = 8.69\pi/6$.}
\label{fig_1}
\end{figure}

\begin{figure}
\includegraphics[width=1\columnwidth]{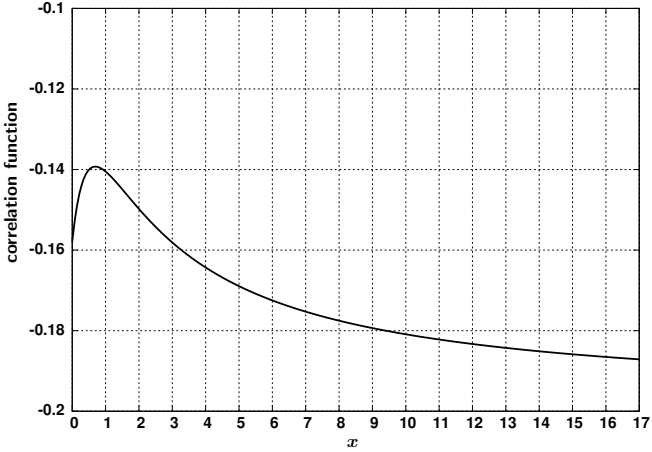}
\caption{The plot shows the dependence of the correlation function
  $\mathcal{C}^{\xi(k,p_k)}_{\textsf{polarization}} 
 (k,\theta;p_k,\tilde{\theta})$
 [Eq.~(\ref{eq:correlation_on_x_alpha})] on $x$ for $\vec{k}$
  orthogonal to $\vec{p}$ ($\alpha=\pi/2$), and $\theta = 5\pi/6$,
  $\tilde{\theta} = 8.69\pi/6$. }
\label{fig_2}
\end{figure}

In the center-of-mass frame [$p=k^\pi=(k^0,-\vec{k})$] the correlation
function (\ref{eq:correl_ksi_polarization_general}) has the form
 \begin{equation}
 \mathcal{C}^{\xi(k,k^\pi)}_{\textsf{polarization}}
 (k,\theta;k^\pi,\tilde{\theta}) =  
 \frac{2(2x+1)^2}{2(2x+1)^2+1} \cos{2(\theta+\tilde{\theta})},
 \label{eq:correl_ksi_polarization_cmf}
 \end{equation}
where $x=(\frac{\vec{k}}{m})^2$. Thus, the correlation function in the
state $\ket{\xi(k,k^\pi)}$ in the center-of-mass frame
[Eq.~(\ref{eq:correl_ksi_polarization_cmf})] is a monotonic
function. However, there exist such configurations of $k$ and $p$ in
which the correlation function
(\ref{eq:correl_ksi_polarization_general}) is not monotonic. For
example, let us assume that $|\vec{p}|=|\vec{k}|$,
$\vec{k}\cdot\vec{p}=|\vec{k}|^2\cos\alpha$. In this configuration
$p=(k^0,\vec{p})\equiv p_k$  and
 \begin{multline}
 \mathcal{C}^{\xi(k,p_k)}_{\textsf{polarization}}
 (k,\theta;p_k,\tilde{\theta}) =  \frac{1}{2(x+1-x\cos\alpha)^2+1}\\
 \times \Big\{  
 \big[ 2x(x+1)(\cos\alpha-1)^2 + \cos^2\alpha + 1 \big]
 \cos(2\theta) \cos(2\tilde{\theta}) \\
 + 2(x+1-x\cos\alpha)(-x+x\cos\alpha+\cos\alpha)
 \sin(2\theta) \sin(2\tilde{\theta})
 \Big\},
 \label{eq:correlation_on_x_alpha}
 \end{multline}
where, as previously, $x=(\frac{\vec{k}}{m})^2$.
We have plotted this function for chosen values of $\theta$ and
$\tilde{\theta}$ on Fig.~\ref{fig_1}. Thus we see that for some values
of $\alpha$ this function is not monotonic function of $x$. As an
example, on Fig.~\ref{fig_2} we have plotted this correlation function
for $\vec{k}\perp\vec{p}$ ($\alpha=\pi/2$).
There are of course also such values of $\theta$ and $\tilde{\theta}$
for which the correlation function (\ref{eq:correlation_on_x_alpha})
is a monotonic function of $x$ for every $\alpha\in(0,\pi\rangle$.

\subsection{Bell-type inequalities}

\begin{figure}
\includegraphics[width=1\columnwidth]{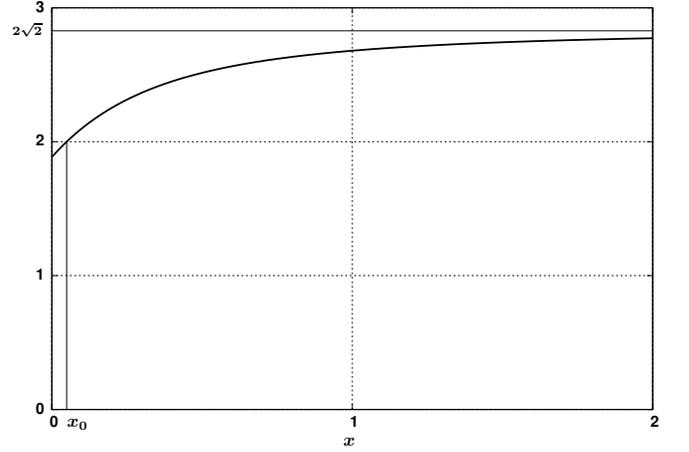}
\caption{The plot shows the dependence of the left side of the
  inequality (\ref{eq:CHSH_polarization_2}) for $\theta_a=0$,
 $\theta_b=\pi/8$, $\theta_c=6\pi/8$, and $\theta_d=3\pi/8$
  on $x$. The inequality (\ref{eq:CHSH_polarization_2}) is violated
  for $x>x_0=-\frac{1}{2}+\frac{1}{2\sqrt{2\sqrt{2}-2}}\approx 0.049$.}
\label{fig_3}
\end{figure}

In this subsection we will show that the degree of violation of
Bell-type inequalities strongly depends on the particle momenta. In
the local realistic theory the CHSH
inequality \cite{cab_CHSH1969,cab_Ballentine1998} 
 \begin{equation}
 |\mathcal{C}_{ab}-\mathcal{C}_{ad}| +
 |\mathcal{C}_{cb}+\mathcal{C}_{cd}| \le 2 
 \label{eq:CHSH_general}
 \end{equation}
should hold. In Eq.~(\ref{eq:CHSH_general}) $\mathcal{C}_{ab}$ denotes
the correlation function of observables with eigenvalues $1$, $0$, and
$-1$, parametrized by $a$ and $b$, and measured by Alice and
Bob. Therefore, the CHSH inequality (\ref{eq:CHSH_general}) in the
local realistic theory should be satisfied also for the polarization
correlation function
(\ref{eq:correl_function_polarization_general}). Let us consider the
CHSH inequality (\ref{eq:CHSH_general}) for the polarization
correlation function in the state $\ket{\xi(k,k^\pi)}$ in the center
of mass frame 
 \begin{multline}
 \big|\mathcal{C}^{\xi(k,k^\pi)}_{\textsf{polarization}}
 (k,\theta_a;k^\pi,\theta_b) - \mathcal{C}^{\xi(k,k^\pi)}_{\textsf{polarization}}
 (k,\theta_a;k^\pi,\theta_d)\big| \\
 + \big|\mathcal{C}^{\xi(k,k^\pi)}_{\textsf{polarization}}
 (k,\theta_c;k^\pi,\theta_b) + \mathcal{C}^{\xi(k,k^\pi)}_{\textsf{polarization}}
 (k,\theta_c;k^\pi,\theta_d)\big| \le 2
 \label{eq:CHSH_polarization_1}
 \end{multline}
Inserting Eq.~(\ref{eq:correl_ksi_polarization_cmf}) into
Eq.~(\ref{eq:CHSH_polarization_1}) we get 
 \begin{multline}
 \frac{2(2x+1)^2}{2(2x+1)^2+1} \Big\{ 
 \big|\cos2(\theta_a+\theta_b) - \cos2(\theta_a+\theta_d)\big| \\
 + \big|\cos2(\theta_c+\theta_b) + \cos2(\theta_c+\theta_d)\big| 
 \Big\} \le 2
 \label{eq:CHSH_polarization_2}
 \end{multline}
The left side of Eq.~(\ref{eq:CHSH_polarization_2}) is largest in the
configuration in which $\big|\cos2(\theta_a+\theta_b) -
\cos2(\theta_a+\theta_d)\big| + \big|\cos2(\theta_c+\theta_b) +
\cos2(\theta_c+\theta_d)\big|$ has maximum 
value. But the analysis of the standard nonrelativistic CHSH
inequality shows that the maksimum value of this quantity is equal to
$2\sqrt{2}$ (we receive this value for e.g.\ $\theta_a=0$,
$\theta_b=\pi/8$, $\theta_c=6\pi/8$, and $\theta_d=3\pi/8$). The
dependence of the left side of inequality
(\ref{eq:CHSH_polarization_2}) in the above configuration is shown in
Fig.~\ref{fig_3}. 
One can see that the degree of violation of this inequality strongly
depends on the particle momentum. Moreover, the left side of the
inequality (\ref{eq:CHSH_polarization_2}) increases with the particle
momentum and reaches the limiting value $2\sqrt{2}$ in the
ultrarelativistic (massless) limit. It should be also noted that in
the nonrelativistic case ($x=0$) the inequality
(\ref{eq:CHSH_polarization_2}) is not violated (the left side of the
inequality (\ref{eq:CHSH_polarization_2}) is equal to
$\frac{4\sqrt{2}}{3}$).

\section{Conclusions}

We have calculated the helicity and linear polarization correlation
functions of two relativistic vector bosons in certain scalar
states. To discuss linear polarization correlations we have defined
states corresponding to the longitudal and transversal polarization of
vector particle. In particular we have found the scalar state which in
the massless limit tends to the two-photon scalar state. We have shown
also that in the massless limit the polarization correlation function
in this scalar state tends to the correlation function in
the scalar two-photon state. We have considered also the CHSH
inequality and found that the degree of violation of this inequality
can increase with the particle momentum. Such a behaviour can be
important in applications of relativistic massive particles in various
quantum information protocols, like e.g.\ quantum cryptography based
on violation of Bell inequalities \cite{cab_Ekert1991}.

\begin{acknowledgments}
The author is grateful to Jakub Rembieli\'nski for his help and
discussion. 
This work has been supported by the University of Lodz.
\end{acknowledgments}


\end{document}